# A simple alteration of the peridynamics correspondence principle to eliminate zero-energy deformation

*Shubhankar Roy Chowdhury[1], Pranesh Roy[1], Debasish Roy[1] and J N Reddy[2,*]*

[1]Computational Mechanics Lab, Department of Civil Engineering, Indian Institute of Science, Bangalore 560012, India

[2]Advanced Computational Mechanics Lab, Department of Mechanical Engineering, Texas A&M University, College Station, Texas 77843-3123

([*]Corresponding author; email: jnreddy@tamu.edu)

*Abstract*

*We look for an enhancement of the correspondence model of peridynamics with a view to eliminating the zero-energy deformation modes. Since the non-local integral definition of the deformation gradient underlies the problem, we initially look for a remedy by introducing a class of localizing corrections to the integral. Since the strategy is found to afford only a reduction, and not complete elimination, of the oscillatory zero-energy deformation, we propose in the sequel an alternative approach based on the notion of sub-horizons. A most useful feature of the last proposal is that the setup, whilst providing the solution with the necessary stability, deviates only marginally from the original correspondence formulation. We also undertake a set of numerical simulations that attest to the remarkable efficacy of the sub-horizon based methodology.*

**Keywords:** Peridynamics; constitutive correspondence; zero-energy modes

## 1. Introduction

Since its introduction in 2000, Peridynamics (PD) (Silling, 2000) has quickly caught the attention of solid mechanicians. An oft-stated advantage with the PD has been its immanent ability to track spontaneously emerging and propagating cracks in solids. In demonstrating this feature, several articles (e.g., Silling and Askari 2005, Silling and Bobaru, 2005) treating static or dynamic fracture of brittle nature have appeared. There is however a serious limitation of the bond-based PD, as originally proposed, in that it is only applicable for isotropic materials with Poisson's ratio 0.25. After nearly a decade, the scope of PD has been extended by the so called state-based PD (Silling *et al.*, 2007), which is applicable to elastic materials with Poisson's ratio in the entire permissible range. Moreover, based on a correspondence principle (Silling *et al.*, 2007), the PD setup is now able to exploit the rich material library in the classical solid mechanics literature, generated by continuous efforts of researchers spanning decades of experimental and theoretical research. By embracing the classical constitutive models, PD offers as an attractive framework to simulate dynamic fracture even for the industrial practitioners. In a similar vein, correspondence-based PD models for problems involving plasticity (Silling *et al.*, 2007), heat conduction (Oterkus *et al.* 2014), micropolar solids (Chowdhury *et al*., 2015), beams (Chowdhury *et al*., 2015), shells (Chowdhury *et al*., 2016), phase field (Roy *et al.* 2017) are now available.

The correspondence framework, however, does not come without a cost. Solutions of PD equations with correspondence materials show instability, e.g. spurious zero-energy modes corrupt the approximated solution field. This becomes more prominent as the influence of nonlocality increases with larger horizon size. Initially it was thought to be an issue with the numerical implementation of the PD and, as early attempts to suppress these spurious modes, additional penalty stiffness was applied leading to a modification of the actual force state (Breitenfeld *et al.* 2014). Several strategies of a similar kind were proposed and applied with varying degrees of success. However, Tupek and Radovitzky (2014) first showed that the source of this instability was not of numerical origin, rather in the very definition of nonlocal deformation gradient used for the correspondence. They proposed an extended constitutive correspondence framework using a generalized Seth-Hill nonlinear strain measure and demonstrated the removal of instability at least for a few one-dimensional problems. However, owing to its strongly nonlinear feature, this correspondence could not create a general appeal and its application remained very restricted. Recently, Silling (2017) has also shown that all correspondence materials as originally proposed in Silling *et al.* (2007) are unstable. He has proposed a scheme to restore stability by introducing additional energy to penalize the difference between local nonuniform deformation and uniform deformation predicted by the nonlocal deformation gradient. This formulation effectively provides a basis for the earlier works wherein additional zero energy force states were added to the original force states to overcome instability in numerical simulations. However, addition of this extra energy to the system appears to be unsatisfactory, a view shared by the authors of Breitzman and Dayal (2018) where they have reformulated the framework of PD with correspondence material. Introducing the notion of bond level nonlocal deformation gradient, they have formulated a stable PD technique. Unlike Silling (2017), no additional penalty energy is used in Breitzman and Dayal (2018). However, along with the new technique of constitutive correspondence, even the equations of motion of PD are significantly altered in this proposition. Certainly, from the point of view of a user, implementation of this new formalism would require extensive modification of the existing codes. We believe this is a weak point for this otherwise interesting work.

In this article, we develop a simple and computationally expedient technique to define the PD correspondence material and associated equations of motion. The new formulation departs only marginally from the original one in Silling *et al.* (2007) whilst imparting the necessary stability to the solution. Our formulation neither introduces additional penalizing energy nor leads to any change in the equations of motion. Thus it allows one to seamlessly incorporate the strategy within already existing PD codes with only an insignificant increase in the computation.

We have organized the rest of the paper as follows. In section 2, we briefly recapitulate the PD equations of motion (EOM), the constitutive correspondence technique and the sources of the zero energy mode instability. Following a passing discussion on the original kinematic correspondence technique that introduces the definition of nonlocal deformation gradient, we propose a corrective strategy to improve this definition in section 3. The corrected nonlocal deformation gradient, when used in the constitutive correspondence, may enhance the stability of the solution of PD equilibrium equations. However, the very reason for which zero-energy modes appear in the original correspondence formulation continues to persist in the corrected strategy too. In section 4, we thus set forth the main proposition - the sub-horizon based PD formulation, and derive all the associated relations, e.g. equations of motion, constitutive relations etc. Following this, in section 5, we demonstrate the effectiveness of the formulation numerically for both one and two dimensional problems. The work is concluded in section 6.

## 2. PD equations of motion, constitutive correspondence and zero energy modes

In this section, we briefly review the non-ordinary state based PD and the framework of constitutive correspondence as proposed in Silling *et al*. (2007). Zero energy modes will also be discussed following Breitenfeld *et al*. (2014).

The referential description of the PD equation of motion (EOM) for a deforming continuum is given by the following integro-differential equation.

$$\rho(\mathbf{x})\ddot{\mathbf{y}}(\mathbf{x},t) = \int_{\mathcal{H}(\mathbf{x})} \left\{ \underline{\mathbf{T}}[\mathbf{x},t]\langle \boldsymbol{\xi} \rangle - \underline{\mathbf{T}}[\mathbf{x}+\boldsymbol{\xi},t]\langle -\boldsymbol{\xi} \rangle \right\} dV' + \mathbf{b}(\mathbf{x},t) \tag{1}$$

Here $\mathbf{x}$ is a material point in the reference configuration $\Omega$, $\rho$ the mass density, $\mathbf{y}(\mathbf{x},t)$ the deformed coordinate of $\mathbf{x}$ at time $t$, $\underline{\mathbf{T}}$ the force vector state and $\mathbf{b}$ the body force density. $\boldsymbol{\xi} = \mathbf{x}' - \mathbf{x}$ is the bond vector between a material point $\mathbf{x}$ and its neighbour $\mathbf{x}'$. $dV'$ denotes the infinitesimal volume measure associated with $\mathbf{x}' = \mathbf{x} + \boldsymbol{\xi}$. The family of neighbouring points considered for a point $\mathbf{x}$ is given by its *horizon* $\mathcal{H}$ defined as $\mathcal{H}(\mathbf{x}) = \{\mathbf{x}' \in \mathbb{R}^3 : |\boldsymbol{\xi} = \mathbf{x}' - \mathbf{x}| < \delta_H\}$, where $\delta_H > 0$ is the radius of the horizon.

Constitutive closure for the EOM is obtained through the constitutive correspondence method. The correspondence relations are given below.

$$\overline{\mathbf{F}} = \left[ \int_{\mathcal{H}(\mathbf{x})} w(|\boldsymbol{\xi}|) \underline{\mathbf{Y}}\langle \boldsymbol{\xi} \rangle \otimes \boldsymbol{\xi} \, dV' \right] \overline{\mathbf{K}}^{-1} \tag{2}$$

$$\overline{\mathbf{K}} = \int_{\mathcal{H}(\mathbf{x})} w(|\boldsymbol{\xi}|) \boldsymbol{\xi} \otimes \boldsymbol{\xi} \, dV' \tag{3}$$

$$\underline{\mathbf{T}} = w \overline{\mathbf{P}} \overline{\mathbf{K}}^{-1} \boldsymbol{\xi} \tag{4}$$

Here, $\underline{\mathbf{Y}}$ is the deformation vector state defined as $\underline{\mathbf{Y}}\langle \boldsymbol{\xi} \rangle = \mathbf{y}(\mathbf{x}') - \mathbf{y}(\mathbf{x})$, $w(|\boldsymbol{\xi}|)$ a scalar valued spherically symmetric influence function, $\overline{\mathbf{K}}$ the shape tensor, $\overline{\mathbf{P}} = \tilde{\mathbf{P}}(\overline{\mathbf{F}})$ the first Piola-Kirchhoff stress tensor obtained from the classical response function $\tilde{\mathbf{P}}$ using the nonlocal deformation gradient $\overline{\mathbf{F}}$ in lieu of its local counterpart.

Although the constitutive correspondence method admits a class of classical constitutive relations to be used with the PD, it also introduces zero energy modes in the solution. This manifests itself in the form of unphysical oscillations and matter interpenetration. As noted in

Breitenfeld *et al.* (2014) and Tupek and Radovitzky (2014), it is due to the weak definition of $\bar{\mathbf{F}}$, originally a gradient field, through an integral that involves spherically symmetric $w$. The upshot is that not every deformation field influences $\bar{\mathbf{F}}$, i.e. there exists a class of deformations leading to the same $\bar{\mathbf{F}}$ and hence the same strain energy density. If $w$ were chosen non-symmetric, which should be reflective of the inherent heterogeneities around a material point, the problem of zero energy might not arise. However, a mathematically congruent choice for non-symmetric $w$ would depend on a complex (perhaps non-Euclidean) geometric description of the microstructural heterogeneities, thereby encrusting the model with considerable complexity. Our present objective is therefore to look for mathematically simpler and computationally more expedient alternatives.

## 3. Nonlocal deformation gradient and correction

The nonlocal deformation gradient is an approximation to (or, depending on the context, a nonlocal generalization of) the local deformation gradient. When interpreted as an approximant, one may find the order of approximation using the definition in equation (2) and Taylor series expansion of $\underline{\mathbf{Y}}\langle \boldsymbol{\xi} \rangle$ as follows:

$$\begin{aligned}
\bar{\mathrm{F}}_{mn} &= \int_{\mathcal{H}} w(|\boldsymbol{\xi}|)\left[\frac{\partial \mathrm{y}_m}{\partial \mathrm{x}_j}\xi_j + \frac{1}{2!}\frac{\partial^2 \mathrm{y}_m}{\partial \mathrm{x}_j \partial \mathrm{x}_k}\xi_j\xi_k + \frac{1}{3!}\frac{\partial^3 \mathrm{y}_m}{\partial \mathrm{x}_j \partial \mathrm{x}_k \partial \mathrm{x}_p}\xi_j\xi_k\xi_p + \cdots\right]\xi_q dV' \bar{\mathrm{K}}_{qn}^{-1} \\
&= \frac{\partial \mathrm{y}_m}{\partial \mathrm{x}_n} + \frac{1}{2!}\frac{\partial^2 \mathrm{y}_m}{\partial \mathrm{x}_j \partial \mathrm{x}_k}\int_{\mathcal{H}} w(|\boldsymbol{\xi}|)\xi_j\xi_k\xi_q dV'\bar{\mathrm{K}}_{qn}^{-1} + \frac{1}{3!}\frac{\partial^3 \mathrm{y}_m}{\partial \mathrm{x}_j \partial \mathrm{x}_k \partial \mathrm{x}_p}\int_{\mathcal{H}} w(|\boldsymbol{\xi}|)\xi_j\xi_k\xi_p\xi_q dV'\bar{\mathrm{K}}_{qn}^{-1} + \cdots \quad (5)\\
&= \mathrm{F}_{mn} + O\left(\delta_H^2\right)
\end{aligned}$$

where $\bar{\mathrm{F}}_{mn}$ refers to the $(m,n)^{\text{th}}$ component of $\bar{\mathbf{F}}$ and $\mathrm{F}_{mn} = \dfrac{\partial \mathrm{y}_m}{\partial \mathrm{x}_n}$ is the $(m,n)^{\text{th}}$ component of the local deformation gradient $\mathbf{F}$. Whilst identifying the order of approximation in equation (5), we have used the following:

$$\int_{\mathcal{H}} w(|\boldsymbol{\xi}|)\xi_j\xi_k\xi_q dV' = 0 \quad (6)$$

In order to increase the order of the accuracy of this approximation, we propose a corrective strategy. Without a loss of generality, we will demonstrate the procedure only for the 1-dimensional case. Rewriting equation (5) for this case, we have:

$$\overline{F} := \int_{-\delta_H}^{\delta_H} w(y'-y)\xi \, dx' \overline{K}^{-1}$$
$$= \frac{\partial y}{\partial x} + \frac{1}{2!}\frac{\partial^2 y}{\partial x^2} \int_{-\delta_H}^{\delta_H} w(|\xi|)\xi^3 dV' \overline{K}^{-1} + \frac{1}{3!}\frac{\partial^3 y}{\partial x^3} \int_{-\delta_H}^{\delta_H} w(|\xi|)\xi^4 dV' \overline{K}^{-1} + \cdots = F + O(\delta_H^2) \quad (7)$$

We introduce a scalar state $w_c$ as:

$$w_c[x]\langle\xi\rangle = w(|\xi|)C(x,\xi) \quad (8)$$

where $C(x,\xi)$ is a correction function written in the following polynomial form.

$$C(x,\xi) = c_0(x) + c_1(x)\xi + c_2(x)\xi^2 + \cdots + c_{N-1}(x)\xi^{N-1} \quad (9)$$

Here $N$ is the index denoting the order of accuracy. Using equations (7) to (9), we may obtain:

$$\int_{-\delta_H}^{\delta_H} w_c(y'-y)\xi \, dx' = \frac{\partial y}{\partial x}(c_0 I_2 + c_1 I_3 + \cdots + c_{N-1} I_{N+1}) + \frac{1}{2!}\frac{\partial^2 y}{\partial x^2}(c_0 I_3 + c_1 I_4 + \cdots + c_{N-1} I_{N+2})$$
$$+ \frac{1}{3!}\frac{\partial^3 y}{\partial x^3}(c_0 I_4 + c_1 I_5 + \cdots + c_{N-1} I_{N+3}) + \cdots + \frac{1}{N!}\frac{\partial^N y}{\partial x^N}(c_0 I_{N+1} + c_1 I_{N+2} + \cdots + c_{N-1} I_{2N})$$
$$+ O(\delta_H^{N+3})$$

(10)

where

$$I_n = \int_{-\delta_H}^{\delta_H} \xi^n dx' \quad \text{for } n \in \{2,3,\cdots,2N\} \quad (11)$$

We choose $c_0, c_1, \cdots, c_{N-1}$ in such a way that the coefficient of $\frac{\partial y}{\partial x}$ in equation (10) is given by

$$c_0 I_2 + c_1 I_3 + \cdots + c_{N-1} I_{N+1} = I_2 \quad (12)$$

and the coefficients of $\frac{\partial^i y}{\partial x^i}$ for $i = 2, 3, \cdots, N$ are zero. In other words, $c_0, c_1, \cdots, c_{N-1}$ are found by solving the following set of linear algebraic equations.

$$\begin{bmatrix} I_2 & I_3 & \cdots & I_{N+1} \\ & I_4 & \cdots & I_{N+2} \\ & & \ddots & \vdots \\ sym & & & I_{2N} \end{bmatrix} \begin{Bmatrix} c_0 \\ c_1 \\ \vdots \\ c_{N-1} \end{Bmatrix} = \begin{Bmatrix} I_2 \\ \{0\}_{(N-1) \times 1} \end{Bmatrix} \quad (13)$$

where the matrix appearing on the left hand side is symmetric, so the symmetric lower-half is denoted by $sym$. Following the usual practice (as in the literature on mesh-free methods, for instance), we call this matrix the moment matrix. With this, equation (10) may be rewritten as:

$$\int_{-\delta_H}^{\delta_H} w_c (y' - y) \xi dx' = \frac{\partial y}{\partial x} I_2 + O(\delta_H^{N+3}) \quad (14)$$

A modified definition for the nonlocal deformation gradient $\overline{F}_c$ may now be introduced as:

$$\overline{F}_c := \int_{-\delta_H}^{\delta_H} w_c (y' - y) \xi dx' I_2^{-1} = \frac{\partial y}{\partial x} + O(\delta_H^N) = F + O(\delta_H^N) \quad (15)$$

It is clear that for $N > 2$, $\overline{F}_c$ is a better approximation of F than $\overline{F}$. Note that, $I_2 = \overline{K}$.

To assess how the correction procedure works in computing the deformation gradient, let us consider a deformation field

$$y = x + a \sin(2\pi k x / L) \quad (16)$$

Choosing $a = 0.9L/(2\pi k)$, $k = 1$, $L = 0.2$ and $\delta_H = 0.5L$, we compute F, $\overline{F}$ and $\overline{F}_c$ for several correction orders, and report them in Figure 1. It is evident that while the conventional PD nonlocal deformation gradient is far away from the local deformation gradient, the modified deformation gradient approaches to the local value with increasing correction order.

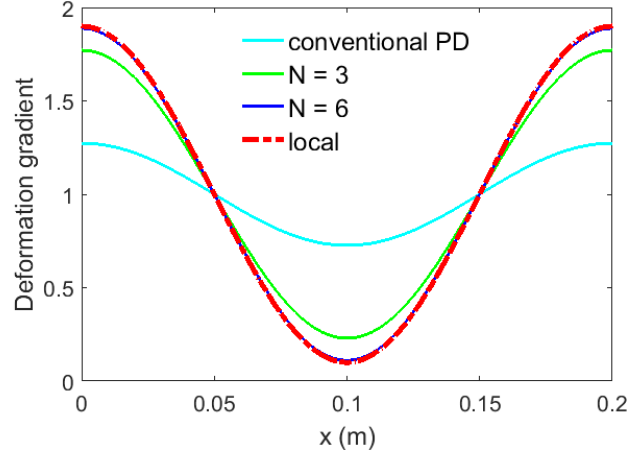

Figure 1: Local deformation gradient and its PD counterpart with and without correction. Correction orders considered are $N = 3$ and $6$.

Even though the correction procedure is demonstrated for the 1-D case, one can extend the same to obtain the corrected nonlocal deformation gradient for 3-D deformation. The correction procedure outlined above is similar to that adopted in mesh-free techniques to derive consistent shape functions (Aluru, 1999). In the context of defining PD differential operators, Madenci *et al*. (2016) have also used a similar strategy.

**Constitutive correspondence with corrected nonlocal deformation gradient**

Writing the classical strain energy density as a function of $\bar{F}_c$, i.e. $\Psi = \tilde{\Psi}(\bar{F}_c)$, its variation may be stated as:

$$\delta\Psi = \left(\mathcal{D}_{\bar{F}_c}\Psi\right)\delta\bar{F}_c = \bar{P}_c\,\delta\bar{F}_c = \bar{P}_c\int_{-\delta_H}^{\delta_H} w_c \delta(y'-y)\xi\,dx'\,I_2^{-1} = \int_{-\delta_H}^{\delta_H} w_c \bar{P}_c\, I_2^{-1}\xi\,\delta(y'-y)\,dx' \qquad (17)$$

Here, $\mathcal{D}_{\bar{F}_c}\Psi = \bar{P}_c = \tilde{P}(\bar{F}_c)$. One can identify the right most term as the variation of the PD strain energy density, $\Psi^{PD} = \tilde{\Psi}^{PD}(y'-y)$:

$$\int_{-\delta_H}^{\delta_H} w_c \bar{P}_c\, I_2^{-1}\xi\,\delta(y'-y)\,dx' = \int_{-\delta_H}^{\delta_H} \underline{T}\,\delta(y'-y)\,dx' = \mathcal{D}_{(y'-y)}\Psi^{PD}\bullet\delta(y'-y) = \delta\Psi^{PD} \qquad (18)$$

where $\underline{T} := w_c \overline{P}_c I_2^{-1} \xi$. From Silling *et al*. (2007), one may note that the new constitutive correspondence is obtainable by replacing the influence function $w$ by the corrected one ($w_c$) and constitution for the first Piola-Kirchhoff stress is derived from $\overline{F}_c$ rather than $\overline{F}$.

In order to evaluate the performance of the proposed strategy, we solve the same problem of axial deformation of a 1-D bar with spatially varying Young's modulus as considered in Breitenfeld *et al*. (2014). For a detailed description of the problem statement refer section 5.1. Figures 2(a) and 2(b) report the solutions of the bar problem with conventional correspondence material for different horizon sizes. As expected, with increasing horizon size the solution gets corrupted with stronger oscillations. Figure 2(b) shows that, for a moderately high horizon size ($\delta_H = 5.1\Delta x$, where $\Delta x$ refers to the mesh size), the oscillations are so strong that even the average trend of deformation is not revealed. Figure 3 offers a contrasting picture by demonstrating the stability brought in through the correction procedure. With increasing correction order, the stability increases. However, from the inset of Figure 3, one also observes that oscillations in the solution, though reduced, persist even after increasing the order of correction. This may be attributed to the fact that, even if correction is introduced, the essential weakness in the definition of the deformation gradient is not eliminated. All the solutions presented in Figure 2 and 3 correspond to a fixed mesh, $\Delta x = L/4000$.

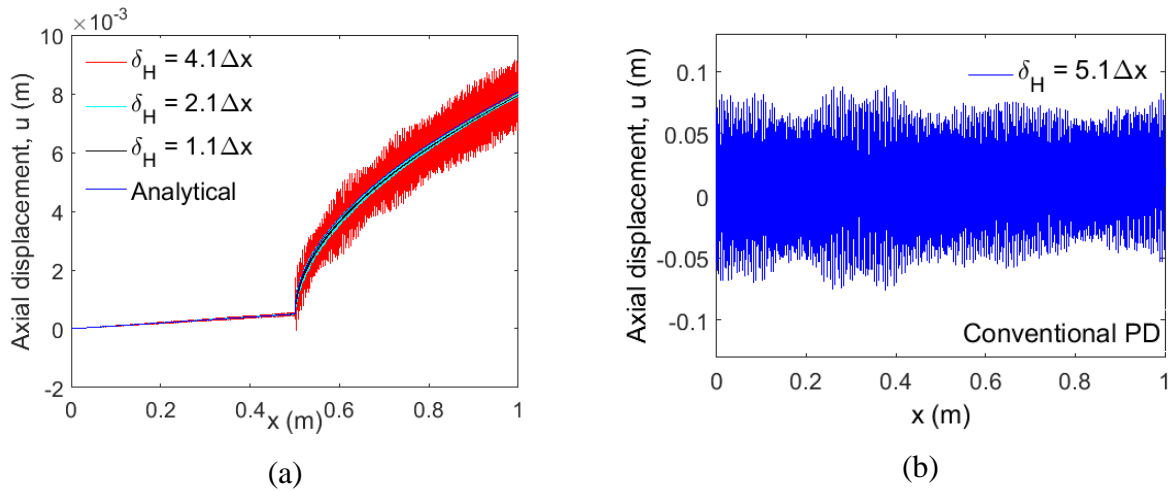

Figure 2: Spatial variation of displacement field: (a) local analytical solution and conventional PD solution with horizon sizes $1.1\Delta x$, $2.1\Delta x$ and $4.1\Delta x$ (b) conventional PD solution with horizon size $5.1\Delta x$

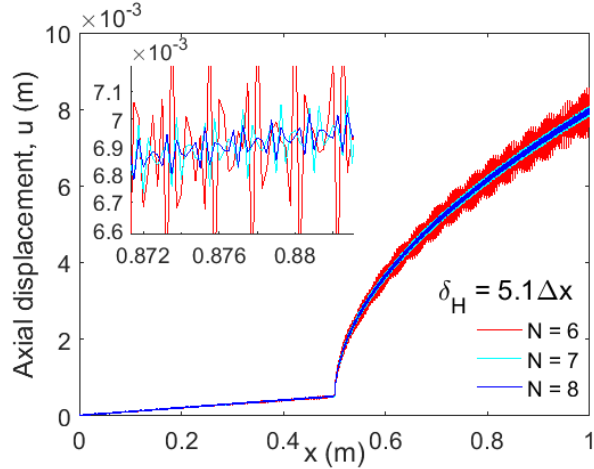

Figure 3: Spatial variation of displacement field obtained through corrected PD for horizon size $5.1\Delta x$. Displayed solutions correspond to correction orders $N = 6, 7, 8$.

## 4. Sub-horizon based PD formulation

Our goal now is to design a strategy which can, in principle, completely eliminate the instability due to zero energy modes in the solutions of PD correspondence models. The reason behind these unphysical results is known to be in the weak definition of the nonlocal deformation gradient which, being an integral expression with a spherically symmetric weight, may remain unchanged for a fairly broad class of deformation fields. In addition, matter interpenetration and sub-horizon collapse may go undetected. Our proposal on the sub-horizon based PD formulation is to overcome these limitations. To accomplish this task, we wish to prevent, with minimally added computational overhead, the cancellation effect in the integral that defines the nonlocal deformation gradient. If such were to be the case, the energetically undetected deformation modes in the old formulation should impart finite energy.

### 4.1. Sub-horizon based nonlocal deformation gradient

Instead of defining the nonlocal deformation gradient through an integration over the entire spherical horizon (see equation 2), we introduce a set of equivalent nonlocal deformation gradients, for each of which the integration is performed over different sub-regions of the horizon (in short sub-horizons). Equivalence follows from the fact that all of these approximate the local deformation gradient at the same material point. However, information on the deformation field they utilize is different for each of them.

At this stage, the choice of the number of such sub-horizons is left arbitrary. Later, we will address, by means of numerical simulations on 1-D and 2-D problems, issues related to an optimal such choice in addition to those on the overall effectiveness of our proposal.

Let the PD horizon $\left(\mathcal{H}\subset\mathbb{R}^n\right)$ be divided into $N_{SH}$ sub-horizons

$$\mathcal{H} = \bigcup_{\alpha=1}^{N_{SH}} \mathcal{H}_\alpha, \tag{19}$$

so that $\mathcal{H}_\alpha$ denotes the $\alpha^{\text{th}}$ sub-horizon. In the following sections, the subscript $\alpha$ will be used for denoting variables pertaining to the sub-horizon $\mathcal{H}_\alpha$. We now introduce the sub-horizon based nonlocal deformation gradients $\left(\bar{\mathbf{F}}_\alpha\right)$ as:

$$\begin{aligned}\bar{\mathbf{F}}_\alpha &= \left[\int_{\mathcal{H}_\alpha} w\left(|\boldsymbol{\xi}|\right) \underline{\mathbf{Y}}\langle\boldsymbol{\xi}\rangle \otimes \boldsymbol{\xi}\, dV'\right]\bar{\mathbf{K}}_\alpha^{-1} \\ &= \left[\int_{\mathcal{H}} \mathcal{I}_\alpha(\boldsymbol{\xi})\, w\left(|\boldsymbol{\xi}|\right) \underline{\mathbf{Y}}\langle\boldsymbol{\xi}\rangle \otimes \boldsymbol{\xi}\, dV'\right]\bar{\mathbf{K}}_\alpha^{-1} = \left[\int_{\mathcal{H}} w_\alpha(\boldsymbol{\xi})\, \underline{\mathbf{Y}}\langle\boldsymbol{\xi}\rangle \otimes \boldsymbol{\xi}\, dV'\right]\bar{\mathbf{K}}_\alpha^{-1}\end{aligned} \tag{20}$$

Here, $\mathcal{I}_\alpha(\boldsymbol{\xi})$ is an indicator function defined as:

$$\mathcal{I}_\alpha(\boldsymbol{\xi}) = \begin{cases} 1 & \text{for } \mathbf{x}' \in \mathcal{H}_\alpha(\mathbf{x}) \\ 0 & \text{otherwise} \end{cases} \tag{21}$$

In equation (20), we have introduced a modified non-spherical influence function $w_\alpha(\boldsymbol{\xi}) := \mathcal{I}_\alpha(\boldsymbol{\xi})\, w\left(|\boldsymbol{\xi}|\right)$. $\bar{\mathbf{K}}_\alpha = \int_{\mathcal{H}} w_\alpha(\boldsymbol{\xi})\, \boldsymbol{\xi} \otimes \boldsymbol{\xi}\, dV'$ denotes the sub-horizon based shape tensor.

### 4.2. Sub-horizon based PD equations of motion

To derive the EOM, we start with Hamilton's principle (though alternative routes, explicitly admitting dissipation, are possible):

$$\int_{t_1}^{t_2} \left[ \int_\Omega \delta \mathcal{L} dV + \int_\Omega \mathbf{f} \cdot \delta \mathbf{y} \, dV \right] dt = 0 \,, \qquad t_1 < t_2 \tag{22}$$

Here, $\mathcal{L}$ and $\mathbf{f}$ respectively denote the Lagrangian density and the body force density. $\mathcal{L}$ is defined in the conventional manner as the difference between kinetic $(\psi_k)$ and potential $(\psi_e)$ energy densities.

$$\mathcal{L} = \psi_k - \psi_e \tag{23}$$

The explicit expression for $\psi_k$ is again the conventional one.

$$\psi_k = \frac{1}{2} \rho \dot{\mathbf{y}} \cdot \dot{\mathbf{y}} \tag{24}$$

Using the complete set of sub-horizon based nonlocal deformation gradients, i.e. $\overline{\mathbf{F}}_\alpha$'s, we need to define the potential energy density $(\psi_e)$. As $\overline{\mathbf{F}}_\alpha$'s are equivalent representations for the local deformation gradient, a natural choice for $\psi_e$ would be to consider an average of the corresponding potential energy densities obtained from them, i.e. $\tilde{\Psi}(\overline{\mathbf{F}}_\alpha) = \Psi_\alpha$ where $\tilde{\Psi}$ is the classical constitutive response function for strain energy density.

$$\psi_e = \frac{1}{N_{SH}} \sum_{\alpha=1}^{N_{SH}} \tilde{\Psi}(\overline{\mathbf{F}}_\alpha) \tag{25}$$

As a linear combination of convex functions with non-negative real coefficients is convex, for a given convex function $\tilde{\Psi}$, the definition above leads to a convex energy function $\psi_e$. Therefore this desired property of an energy functional is retained upon modification of definition. Using equations (23) to (25), we write the variation of $\mathcal{L}$ as:

$$\begin{aligned} \delta \mathcal{L} &= \delta \psi_k - \delta \psi_e \\ &= \mathcal{D}_{\dot{\mathbf{u}}} \psi_k \cdot \delta \dot{\mathbf{y}} - \frac{1}{N_{SH}} \sum_{\alpha=1}^{N_{SH}} \mathcal{D}_{\overline{\mathbf{F}}_\alpha} \Psi_\alpha : \delta \overline{\mathbf{F}}_\alpha \\ &= \mathbf{I} \cdot \delta \dot{\mathbf{y}} - \frac{1}{N_{SH}} \sum_{\alpha=1}^{N_{SH}} \overline{\mathbf{P}}_\alpha : \delta \overline{\mathbf{F}}_\alpha \end{aligned} \tag{26}$$

In the last step, we have used the following abbreviations.

$$\mathbf{I} := \mathcal{D}_{\dot{\mathbf{u}}}\psi_k \qquad \bar{\mathbf{P}}_\alpha := \mathcal{D}_{\bar{\mathbf{F}}_\alpha}\Psi_\alpha \tag{27}$$

Substituting $\delta\mathcal{L}$ from equation (26) in the Hamilton's principle (equation 22), we obtain:

$$\int_{t_1}^{t_2}\left[\int_\Omega\left(\mathbf{I}\cdot\delta\dot{\mathbf{y}}-\frac{1}{N_{SH}}\sum_{\alpha=1}^{N_{SH}}\bar{\mathbf{P}}_\alpha:\delta\bar{\mathbf{F}}_\alpha\right)dV+\int_\Omega\mathbf{f}\cdot\delta\mathbf{y}\,dV\right]dt=0 \tag{28}$$

We now consider each term separately and simplify them. The first term can be recast as:

$$\int_{t_1}^{t_2}\int_\Omega\mathbf{I}\cdot\delta\dot{\mathbf{y}}\,dVdt=\int_{t_1}^{t_2}\int_\Omega\mathbf{I}\cdot\frac{D}{Dt}(\delta\mathbf{y})\,dVdt=\int_\Omega\mathbf{I}\cdot\delta\mathbf{y}\,dV\Big|_{t_1}^{t_2}-\int_{t_1}^{t_2}\int_\Omega\frac{D\mathbf{I}}{Dt}\cdot\delta\mathbf{u}\,dVdt$$
$$=-\int_{t_1}^{t_2}\int_\Omega\frac{D\mathbf{I}}{Dt}\cdot\delta\mathbf{y}\,dVdt=-\int_{t_1}^{t_2}\int_\Omega\frac{D}{Dt}(\rho\dot{\mathbf{y}})\cdot\delta\mathbf{y}\,dVdt \tag{29}$$

Similarly, the second term takes the following form.

$$\int_{t_1}^{t_2}\int_\Omega\frac{1}{N_{SH}}\sum_{\alpha=1}^{N_{SH}}\bar{\mathbf{P}}_\alpha:\delta\bar{\mathbf{F}}_\alpha dVdt=\int_{t_1}^{t_2}\int_\Omega\frac{1}{N_{SH}}\sum_{\alpha=1}^{N_{SH}}\bar{\mathbf{P}}_\alpha:\left[\int_\mathcal{H}w_\alpha\delta\underline{\mathbf{Y}}\langle\xi\rangle\otimes\xi dV'\right]\bar{\mathbf{K}}_\alpha^{-1}dVdt$$
$$=\int_{t_1}^{t_2}\int_\Omega\frac{1}{N_{SH}}\sum_{\alpha=1}^{N_{SH}}\bar{\mathbf{P}}_\alpha:\left[\int_\Omega w_\alpha\delta(\mathbf{y}'-\mathbf{y})\otimes\xi\right]\bar{\mathbf{K}}_\alpha^{-1}dV'dVdt$$
$$=\int_{t_1}^{t_2}\int_\Omega\int_\Omega\frac{1}{N_{SH}}\sum_{\alpha=1}^{N_{SH}}w_\alpha'\bar{\mathbf{P}}_\alpha'\bar{\mathbf{K}}_\alpha^{-1'}\xi'\cdot\delta\mathbf{y}dV'dVdt-\int_{t_1}^{t_2}\int_\Omega\int_\Omega\frac{1}{N_{SH}}\sum_{\alpha=1}^{N_{SH}}w_\alpha\bar{\mathbf{P}}_\alpha\bar{\mathbf{K}}_\alpha^{-1}\xi\cdot\delta\mathbf{y}dV'dVdt \tag{30}$$
$$=-\int_{t_1}^{t_2}\int_\Omega\{\underline{\mathbf{T}}[\mathbf{x}]\langle\xi\rangle-\underline{\mathbf{T}}[\mathbf{x}']\langle-\xi\rangle\}\cdot\delta\mathbf{y}dV'dt$$

In the last step (equation 30), we have introduced the notation $\underline{\mathbf{T}}[\mathbf{x}]\langle\xi\rangle$ which is defined as:

$$\underline{\mathbf{T}}[\mathbf{x}]\langle\xi\rangle:=\frac{1}{N_{SH}}\sum_{\alpha=1}^{N_{SH}}w_\alpha\bar{\mathbf{P}}_\alpha\bar{\mathbf{K}}_\alpha^{-1}\xi \tag{31}$$

Substituting equations (29) and (30), in equation (28), we obtain:

$$-\int_{t_1}^{t_2}\int_\Omega\frac{d}{dt}(\rho\dot{\mathbf{y}})\cdot\delta\mathbf{y}\,dVdt+\int_{t_1}^{t_2}\int_\Omega\int_\Omega(\underline{\mathbf{T}}-\underline{\mathbf{T}}')\cdot\delta\mathbf{y}\,dV'dVdt+\int_{t_1}^{t_2}\int_\Omega\mathbf{f}\cdot\delta\mathbf{y}\,dVdt=0 \tag{32}$$

where, for brevity, we have denoted $\underline{\mathbf{T}}[\mathbf{x}]\langle\xi\rangle$ by $\underline{\mathbf{T}}$ and $\underline{\mathbf{T}}[\mathbf{x}']\langle-\xi\rangle$ by $\underline{\mathbf{T}}'$ respectively.

Since equation (32) is valid for any kinematically admissible variation $\delta \mathbf{y}$, the following must hold.

$$\int_\Omega \{\underline{\mathbf{T}} - \underline{\mathbf{T}}'\} dV' + \mathbf{f} = \frac{D}{Dt}(\rho \dot{\mathbf{y}}) \tag{33}$$

This is the Euler-Lagrange equation and represents the PD EOM. Since $\underline{\mathbf{T}} - \underline{\mathbf{T}}'$ vanishes beyond the horizon $\mathcal{H}$, equation (33) may also be written as:

$$\int_\mathcal{H} \{\underline{\mathbf{T}} - \underline{\mathbf{T}}'\} dV' + \mathbf{f} = \frac{D}{Dt}(\rho \dot{\mathbf{y}}) \tag{34}$$

### 4.3. Verification for global balances of linear and angular momenta

Integrating both sides of the equation (33) over $\Omega$, we obtain:

$$\int_\Omega \int_\Omega \{\underline{\mathbf{T}} - \underline{\mathbf{T}}'\} dV' dV + \int_\Omega \mathbf{f}\, dV = \int_\Omega \frac{D}{Dt}(\rho \dot{\mathbf{y}}) dV \tag{35}$$

Exploiting the anti-symmetry of the $\underline{\mathbf{T}} - \underline{\mathbf{T}}'$, the first term of the left hand side of equation (35) may be shown to be zero. This immediately yields the global balance of linear momentum, i.e.,

$$\int_\Omega \mathbf{f}\, dV = \int_\Omega \frac{D}{Dt}(\rho \dot{\mathbf{y}}) dV \tag{36}$$

To verify the global balance of angular momentum, we evaluate the following:

$$\int_\Omega \mathbf{y} \times \left(\frac{D}{Dt}(\rho \dot{\mathbf{y}}) - \mathbf{f}\right) dV = \int_\Omega \int_\Omega \underline{\mathbf{T}} \times \underline{\mathbf{Y}} dV' dV \tag{37}$$

where the right hand side is obtained upon using equation (33) and change of variables $\mathbf{x} \leftrightarrow \mathbf{x}'$ (see Silling et al., 2007). Substituting $\underline{\mathbf{T}}$ from equation (31), we may write:

$$\int_\mathcal{H} \underline{\mathbf{T}} \times \underline{\mathbf{Y}} dV' = \int_\mathcal{H} \frac{1}{N_{SH}} \sum_{\alpha=1}^{N_{SH}} w_\alpha \bar{\mathbf{P}}_\alpha \bar{\mathbf{K}}_\alpha^{-1} \boldsymbol{\xi} \times \underline{\mathbf{Y}} dV' = \int_\mathcal{H} \frac{1}{N_{SH}} \sum_{\alpha=1}^{N_{SH}} w_\alpha J_\alpha \bar{\boldsymbol{\sigma}}_\alpha \mathbf{F}_\alpha^{-T} \bar{\mathbf{K}}_\alpha^{-1} \boldsymbol{\xi} \times \underline{\mathbf{Y}} dV' \tag{38}$$

where $\bar{\boldsymbol{\sigma}}_\alpha = (J_\alpha)^{-1} \mathbf{P}_\alpha \mathbf{F}_\alpha^T$ is the nonlocal symmetric Cauchy stress.

Writing equation (38) in the indicial notation, one may conclude the following:

$$\begin{aligned}\left(\int_{\mathcal{H}} \underline{\mathbf{T}} \times \underline{\mathbf{Y}} dV'\right)_i &= \int_{\mathcal{H}} \frac{1}{N_{SH}} \sum_{\alpha=1}^{N_{SH}} \varepsilon_{ijk} \left(w_\alpha J_\alpha \bar{\boldsymbol{\sigma}}_\alpha \mathbf{F}_\alpha^{-T} \bar{\mathbf{K}}_\alpha^{-1} \boldsymbol{\xi}\right)_j \mathbf{Y}_k \, dV' \\ &= \frac{1}{N_{SH}} \sum_{\alpha=1}^{N_{SH}} \varepsilon_{ijk} J_\alpha (\bar{\boldsymbol{\sigma}}_\alpha)_{jp} (\mathbf{F}_\alpha^{-T})_{pq} (\bar{\mathbf{K}}_\alpha^{-1})_{qm} \int_{\mathcal{H}} w_\alpha \xi_m \mathbf{Y}_k dV' \\ &= \frac{1}{N_{SH}} \sum_{\alpha=1}^{N_{SH}} \varepsilon_{ijk} J_\alpha (\bar{\boldsymbol{\sigma}}_\alpha)_{jp} (\mathbf{F}_\alpha^{-T})_{pq} (\mathbf{F}_\alpha)_{kq} = \frac{1}{N_{SH}} \sum_{\alpha=1}^{N_{SH}} \varepsilon_{ijk} J_\alpha (\bar{\boldsymbol{\sigma}}_\alpha)_{jp} \delta_{kp} \\ &= \frac{1}{N_{SH}} \sum_{\alpha=1}^{N_{SH}} \varepsilon_{ijk} J_\alpha (\bar{\boldsymbol{\sigma}}_\alpha)_{jk} = 0\end{aligned} \quad (39)$$

Here the final step follows from the symmetry of $\bar{\boldsymbol{\sigma}}_\alpha$. From equation (37), one readily arrives at:

$$\int_\Omega \mathbf{y} \times \left(\frac{D}{Dt}(\rho \dot{\mathbf{y}}) - \mathbf{f}\right) dV = \mathbf{0} \quad (40)$$

proving the global balance of angular momentum.

## 5. Numerical assessment of performance

We carry out a set of numerical simulations for 1D and 2D problems in linear elasticity. All the problems are solved using both the conventional correspondence framework of PD and the sub-horizon based one.

### 5.1. Static deformation of 1D 'singular' bar (Breitenfeld *et al*. 2014)

The present illustration concerns a bar of spatially varying Young's modulus fixed at one end and pulled by a traction of $10^4 \text{ N/m}^2$ at the other. The classical constitutive equation of the bar is of the following form.

$$\sigma = E(F-1) \quad (41)$$

where $\sigma$ denotes stress, F the deformation gradient and $E$ Young's modulus whose spatial variation is given as:

$$E = \begin{cases} E_0 & x \leq L/2 \\ E_0\alpha\left(\alpha + \dfrac{0.01}{2\sqrt{x-L/2}}\right)^{-1} & x > L/2 \end{cases} \tag{42}$$

For PD simulations with conventional and sub-horizon based correspondence relations, the PD force state in the bar takes the following respective forms.

$$\underline{T} = wE\left(\overline{F}-1\right)\overline{K}^{-1}\xi \quad \text{and} \quad \underline{T} = \sum_{\alpha=1}^{N_{SH}} w_\alpha E\left(\overline{F}_\alpha - 1\right)\overline{K}_\alpha^{-1}\xi \tag{43}$$

For numerical simulation, consider the following geometric and material properties.

Length $L = 1$ m, $E_0 = 10^7$ N/m$^2$ and $\alpha = 0.001$.

In Figure 4, we report solutions through the sub-horizon based framework with different horizon sizes. The adopted mesh size is $\Delta x = L/4000$. Number of sub-horizons used is 2. The two sub-horizons of a point x is defined as:

$$\mathcal{H}_1(x) = \{x': -\delta_H \leq x' - x \leq 0\} \quad \text{and} \quad \mathcal{H}_2(x) = \{x': 0 \leq x' - x \leq \delta_H\}$$

Unlike solutions obtained using conventional and corrected influence function based PD (Figures 2 and 3) in both of which the presence of zero energy modes manifests through oscillations, sub-horizon based PD results (in Figure 4) are entirely oscillation free even when we have used larger horizon size $\delta_H = 10.1\Delta x$. One observes that, with increasing horizon size, the magnitude of deformation reduces marginally. This is expected because of the stiffness enhancement due to nonlocal effect.

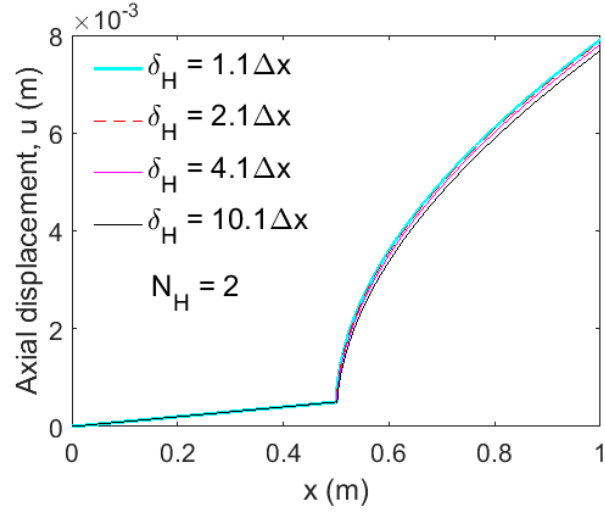

Figure 4: Spatial variation of displacement field through sub-horizon based PD for different horizon sizes; number of sub-horizons considered is 2.

For an assessment of the effect of increasing number of sub-horizons on the computed displacement field, we now consider four sub-horizons as defined below:

$$\mathcal{H}_1(x) = \{x': -\delta_H \leq x' - x \leq -0.5\delta_H\}, \quad \mathcal{H}_2(x) = \{x': -0.5\delta_H \leq x' - x \leq 0\},$$
$$\mathcal{H}_3(x) = \{x': 0 \leq x' - x \leq 0.5\delta_H\} \text{ and } \mathcal{H}_4(x) = \{x': 0.5\delta_H \leq x' - x \leq \delta_H\}$$

Choosing the mesh and horizon sizes as $\Delta x = L/4000$ and $\delta_H = 10.1\Delta x$ respectively, we solve the bar problem with two and four sub-horizons and report the solutions in Figure 5. Observe that both the cases lead to exactly the same solution. This establishes the adequacy (at least for the 1D case) of a minimally split horizon to ensure elimination of zero energy modes.

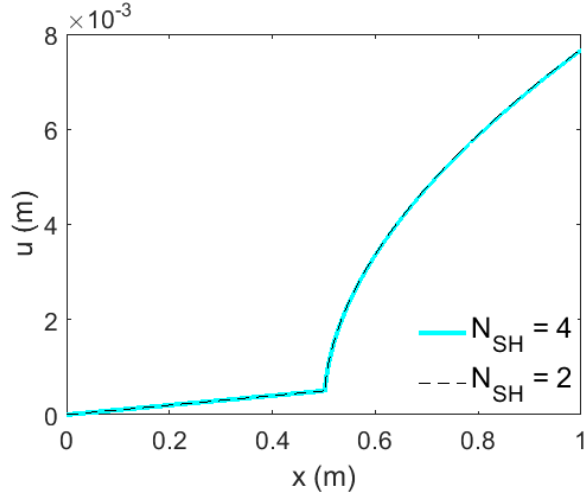

Figure 5: Spatial variation of displacement field obtained through sub-horizon based PD for horizon size $\delta_H = 10.1\Delta x$ with different numbers of sub-horizons.

In numerical implementation of PD, it is possible to consider different types of convergences – (1) convergence with refinement of mesh for a fixed horizon size; (2) convergence with mesh refinement along with diminishing horizon size. Case (2) recovers the local solution in the limit as $\delta_H \to 0$ and $\Delta x \to 0$. In Figure 6, this local solution obtained through sub-horizon based PD is reported and compared with the available analytical solution. Convergence type 1 leads to the solution of a truly nonlocal problem. While in the conventional PD correspondence, this type of convergence is unattainable due to the emergence of zero energy mode instability, the sub-horizon based PD correspondence does indeed attain a convergent solution as shown in Figure 6.

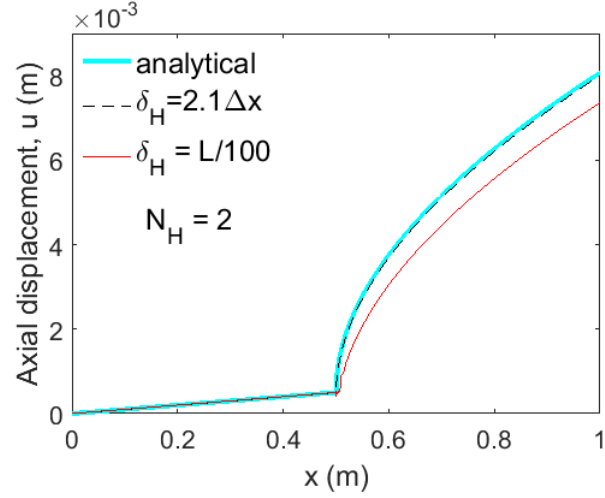

Figure 6: Spatial variation of displacement field through sub-horizon based PD. Local solution obtained with diminishing horizon size and nonlocal solution for a fixed horizon size $\delta_H = 0.01L$ are reported. Number of sub-horizons used is 2.

### 5.2. Static deformation of 2D plate with hole

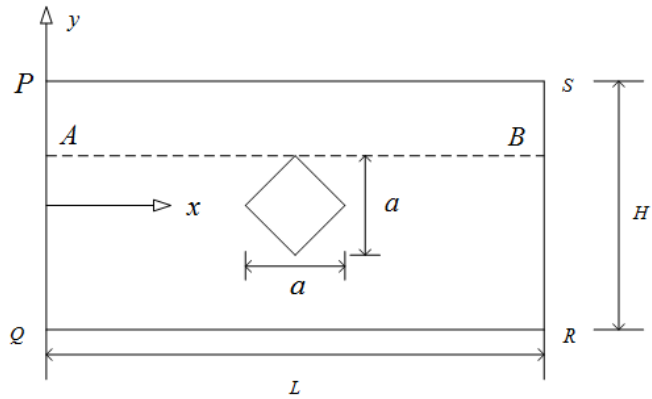

Figure 7: Geometry of the plate with hole and applied boundary conditions

A 2D plate (length $L = 0.15\,\text{m}$, width $H = 0.075\,\text{m}$) with a diamond shaped hole ($a = 0.0375\,\text{m}$) under in-plane tension is investigated. The plate is basically loaded applying displacement boundary conditions at the edges $PQ$ and $SR$ as $u_x|_{PQ} = -10^{-8}\,\text{m}$ and $u_x|_{SR} = 10^{-8}\,\text{m}$. The large deformation gradient introduced by the hole is known to trigger severe zero energy mode instability in the solution when solved through the usual PD correspondence method (Silling,

2017). A hole with sharp corners, as considered presently, is expected to display even severer instability. For the purpose of simulation, the small deformation setting and plane strain conditions are adopted. The material is linear elastic with Young's modulus and Poisson's ratio being 200 GPa and 0.25 respectively.

In Figure 8, we report the x-component of displacement fields based on both conventional and sub-horizon based PD correspondence methods. While the solution via the conventional method shows spurious oscillations (Figure 8a), our sub-horizon based proposal obtains oscillation-free solution (Figure 8b). To make this demonstration more conspicuous, we consider displacements along the line AB (Figure 7) and report them in Figure 9. For this problem, the number of sub-horizons for a point $\mathbf{x}$ is four and their definitions are given below:

$$\mathcal{H}_1(\mathbf{x}) := \{\mathbf{x}': x_1' - x_1 > 0,\ x_2' - x_2 > 0 \text{ and } |\mathbf{x}' - \mathbf{x}| \leq \delta_H\},$$
$$\mathcal{H}_2(\mathbf{x}) := \{\mathbf{x}': x_1' - x_1 < 0,\ x_2' - x_2 > 0 \text{ and } |\mathbf{x}' - \mathbf{x}| \leq \delta_H\},$$
$$\mathcal{H}_3(\mathbf{x}) := \{\mathbf{x}': x_1' - x_1 < 0,\ x_2' - x_2 < 0 \text{ and } |\mathbf{x}' - \mathbf{x}| \leq \delta_H\},$$
$$\mathcal{H}_4(\mathbf{x}) := \{\mathbf{x}': x_1' - x_1 > 0,\ x_2' - x_2 < 0 \text{ and } |\mathbf{x}' - \mathbf{x}| \leq \delta_H\}$$

$(x_1, x_2)$ are the coordinates of the point $\mathbf{x}$. Similar definitions hold for the coordinates of $\mathbf{x}'$.

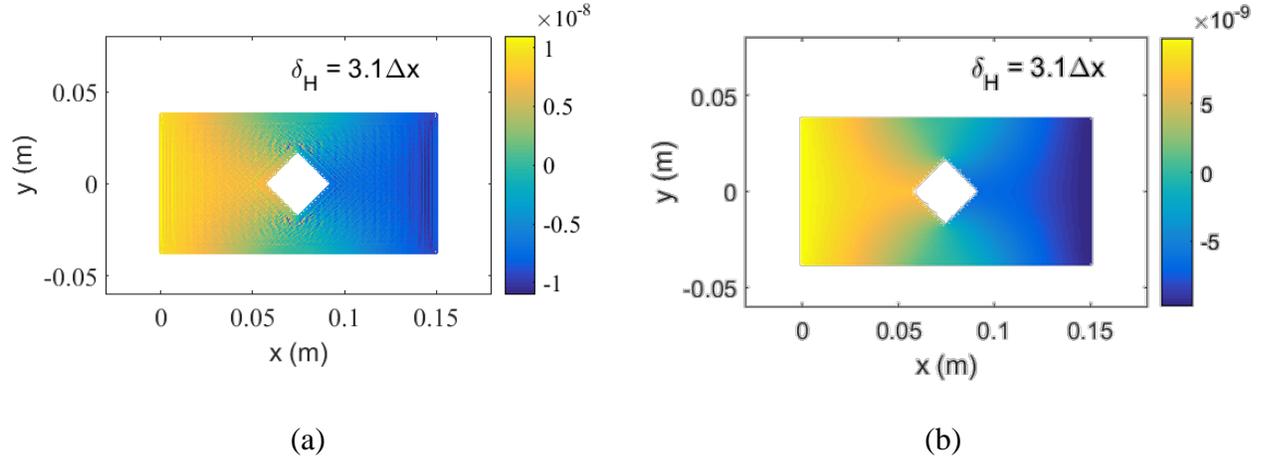

(a)          (b)

Figure 8: x-component of displacement obtained through (a) conventional PD correspondence (b) sub-horizon based PD correspondence. Mesh and horizon sizes are $\Delta x = \Delta y = L/160$ and $\delta_H = 3.1\sqrt{\Delta x^2 + \Delta y^2}$ respectively.

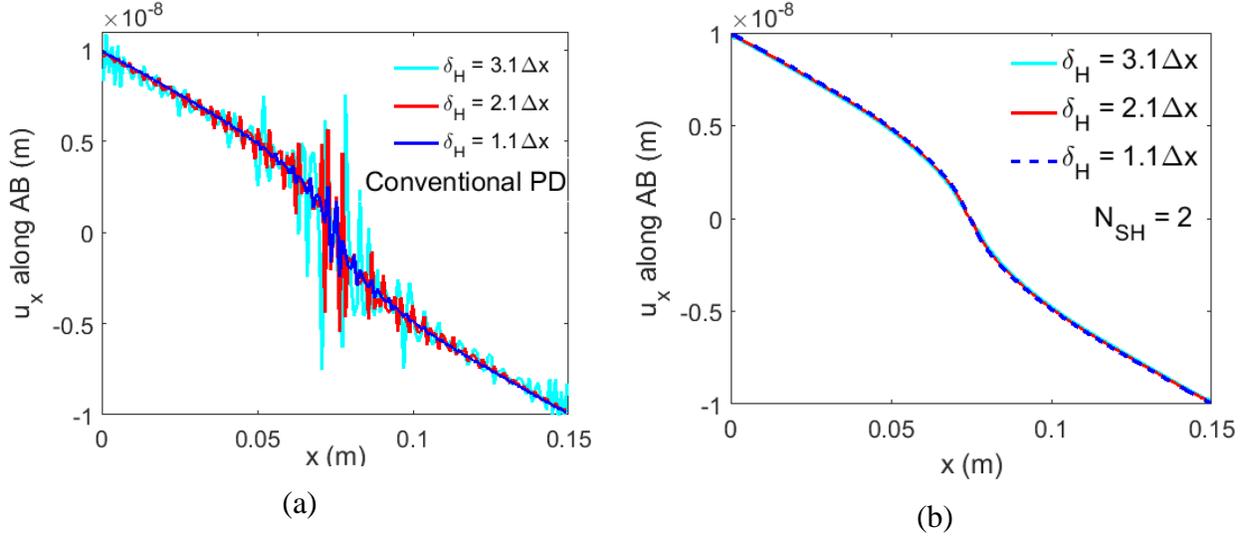

Figure 9: x-component of displacement along AB obtained through (a) conventional PD correspondence (b) sub-horizon based PD correspondence. Mesh and horizon sizes are $\Delta x = \Delta y = L/160$ and $\delta_H / \sqrt{\Delta x^2 + \Delta y^2} = 1.1, 2.1, 3.1$ respectively.

### 5.3. Spectrum in 1D

We impose a discrete axial displacement field defined by:

$$u(x_i) = a \sin(k x_i / L)$$

on a discretized bar of length $L = 1\,\text{m}$ and Young's modulus $E = 1\,\text{N}/\text{m}^2$. $x_i$ denotes the center of the $i$-th segment of size $\Delta x$. $k/L$ refers to the wave number and $a$ the amplitude of the imposed displacement field. Define an energy like measure associated with this displacement as:

$$e(k) := \sum_i u(x_i) \sum_j \left( \underline{T}[x_j]\langle x_i - x_j \rangle - \underline{T}[x_i]\langle x_j - x_i \rangle \right) \Delta x$$

In Figure 10, we have plotted $e$ vs. $k$ where the force state $\underline{T}$ is computed using the conventional PD methodology with horizon size $\delta_H = L/100$. Similar plots are furnished in Figures 11a and 11b, where $\underline{T}$'s are computed respectively from two and four sub-horizon based PD formulations for the same horizon size as above. Three sets of result corresponding to different mesh sizes ($L/\Delta x = 1000, 2000, 4000$) are presented for each case.

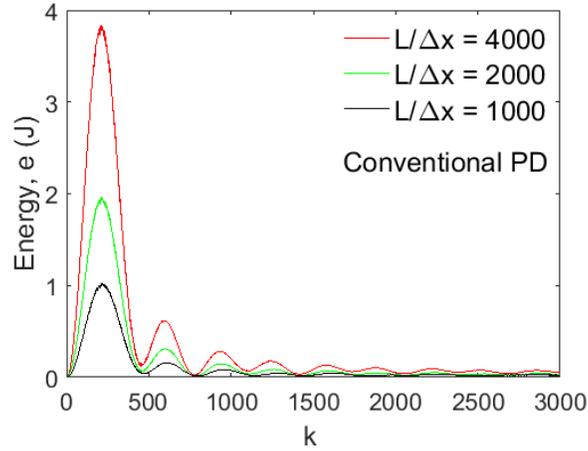

Figure 10: Frequency spectrum of conventional PD correspondence model

From Figure 10, one readily notes that the energy content $e$ for large values of $k$ is close to zero. In fact, one easily verifies that $e \to 0$ as $k \to \infty$. This exercise is a pointer to the existence of zero energy modes (displacement field with $e = 0$) in the conventional PD correspondence model. However, for the sub-horizon based models, the energy content $e$ is always finite irrespective of the value of $k$ (see Figures 11a and 11b). Thus, with the proposed sub-horizon based models, every deformation fields imparts energy, so that no zero energy modes exist. Further, for sub-horizon based models, one may see that, with finer discretization, the value of $e$ increases for all $k$. However, such an increase is practically absent in the conventional PD correspondence model for large value of $k$.

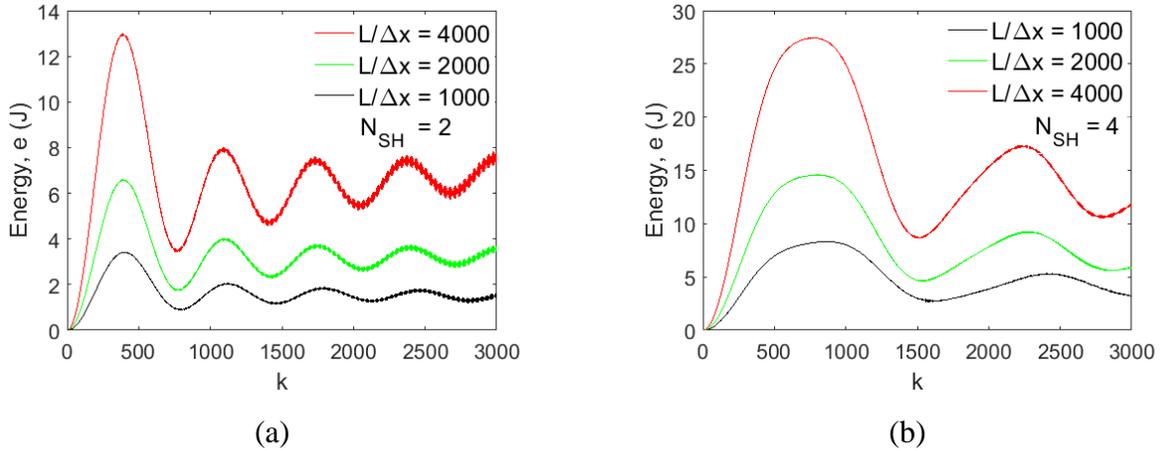

(a)                      (b)

Figure 11: Frequency spectrum of sub-horizon based PD correspondence model: (a) with two sub-horizons and (b) with four sub-horizons

## 6. Conclusions

Our exploration for a conclusive yet simple fix to the zero-energy deformation problem has ended up in a new sub-horizon based reformulation for the correspondence framework of non-ordinary state based peridynamics. The new proposal completely eliminates the zero energy instability that besets conventional formulations, whilst demanding only a minor alteration in the framework of the existing PD correspondence principle. The implication of this work is immediately evident – with an uninvolved modification of standard PD formulations (as available in the existing codes, for instance) and with very little additional demand on computation, zero-energy free solutions may now be obtained.